\documentclass[prl,twocolumn,superscriptaddress]{revtex4}
\usepackage[latin1]{inputenc}

\usepackage{graphicx}

\def \beq {\begin{equation}}
\def \eeq {\end{equation}}
\def \beqa {\begin{eqnarray}}
\def \eeqa {\end{eqnarray}}

\begin{document}

\title{{\sc Generic Finite Size Enhancement of Pairing \\in
Mesoscopic Fermi Systems}}
\author{M. Farine}
\affiliation{Consulat G\'en\'eral de France \`a Canton, 339 Huan Shi Dong 
Lu, 510098 Guangzhou, Canton, China}
\author{F. W. J. Hekking}
\affiliation{Laboratoire de Physique et Mod\'elisation des Milieux
Condens\'es, CNRS and Universit\'e Joseph Fourier, Maison des 
Magist\`eres, 
B.P. 166,
38042 Grenoble Cedex 9, France}
\author{P. Schuck}
\affiliation{Institut de Physique Nucl\'{e}aire, IN{\sl 2}P{\sl 3} -- CNRS,
Universit\'{e} Paris -- Sud, F-91406 Orsay-C\'{e}dex, France}               
\affiliation{Laboratoire de Physique et Mod\'elisation des Milieux
Condens\'es, CNRS and Universit\'e Joseph Fourier, Maison des 
Magist\`eres, 
B.P. 166,
38042 Grenoble Cedex 9, France}
\author{X. Vi\~nas}
\affiliation{Departament d'Estructura i Constituents de la Mat\`{e}ria,
Facultat de F\'{\i}sica, Universitat de Barcelona, Diagonal 647, E-08028
Barcelona, Spain}

\begin{abstract}
The finite size dependent enhancement of pairing in mesoscopic Fermi systems
is studied under the assumption that the BCS approach is valid and that the
two body force is size independent. Different systems are investigated such
as superconducting metallic grains and films as well as atomic nuclei. It is
shown that the finite size enhancement of pairing in these systems is in 
part due to the presence of a surface which accounts quite well for the 
data of nuclei and explains a good fraction of the enhancement in Al grains.
\end{abstract}

\maketitle

%

Since long it is a well-known fact that in certain finite size Fermi 
systems the
gap is increased substantially from its bulk value.\ Such systems are, for
instance, ultra small superconducting metallic grains, of great present 
actuality \cite{BR96,RB97,BD99}, and thin films 
\cite{MG70,CA68,DFGGO73,DGGI73} 
but also superfluid atomic nuclei \cite{BM69,RS80}. 
There have been theoretical studies in the past on the size dependence of
pairing in the abovementioned systems \cite{TB63a,E67,P68,HBGSS02}. To 
our
knowledge for the
condensed matter systems no satisfying explanation has been found \cite{TB63a}
whereas for
the nuclear systems large scale Hartree-Fock-Bogolioubov (HFB) calculations
for
nuclei have recently somewhat clarified the situation \cite{HBGSS02}.

In this investigation we will set a rather limiting frame: we assume that
BCS\ theory is valid and that the pairing force v$\left(
\mbox{\boldmath$r$}%
\right) $ is size independent.\ These are, of course, very severe
restrictions, and obviously, other size dependent features may be present 
in reality. 
 We will consider simplified systems: First we study
metallic grains and films in a hard wall potential using the standard
schematic constant matrix element approximation with an adjustable strength
\ parameter and a cut-off given by the Debye frequency.
It will be shown that this model acounts for a good fraction of
 the experimental size dependence.
Second we apply the previously developed pocket formula to the mass number
dependence of nuclear gaps.
We will see that our simple theory describes the mass number 
($A$) dependence of
nuclear pairing quite well. In all cases only the spin singlet channel shall
be considered.

Let us first present our general approach.\ As already mentioned, we want to
base our consideration on the validity of BCS\ theory.\ In finite systems
the gap equation can therefore be written in the standard form \cite{RS80},
where the states $|n\rangle $ are the eigenvectors of the single particle
Hamiltonian $h=p^{2}/2m^{\ast }+\;V\left( r\right) $ with $V\left( r\right)
$
the (phenomenological) single particle potential and $m^{\ast }=m^{\ast
}\left( r\right) $ the effective mass:
\begin{equation}  \label{form1}
\Delta _{n}=-\sum_{n^{\prime}} {\langle n\ \bar{n}\ |\ v\ |\ n}^{\prime }{\
\bar{n}}^{\prime }\rangle {\ }{\Delta _{n^{\prime }}}/{2 E_{n^{\prime}}}
\end{equation}
In (\ref{form1}) $E_ {n}$ are the quasi-particle energies 
with $E^2_{n}=
\left(
\epsilon _{n}-\mu \right) ^{2}+\Delta _{n}^{2}$ and 
the single particles energies  
$\epsilon _{n}$ are the eigenvalues of $h$, i.e., $h|n\rangle =\epsilon
_{n}|n\rangle$, the pairing matrix element $\langle n\bar{n}|$v$|n^{\prime
}\bar{n}^{\prime }\rangle $ contains the time reversed states $|\overline{n}
\rangle $, and the chemical potential $\mu $ for finite systems is
determined by the "particle number ($N$) condition": $N= \sum_n \frac{1}{2}
\big(1 - {(\epsilon_{n} -\mu )}/{E_{n}}\big)$.
This model, though quite schematic, will allow us
to develop the essential features of the size dependence of pairing.\ One
further important hypothesis, as already mentioned, is that the pairing
force from which the matrix elements in (\ref{form1}) are constructed, does
itself not depend on the size of the system.\ Still the matrix elements, via
the wavefunctions, will be size-dependent.\ One guesses that the other
important sources of mass number dependence in (\ref{form1}) are the single
particle spectrum, respectively the level density $g\left( \epsilon \right)
=\sum_{n}\delta (\epsilon -\varepsilon _{n})$, and the chemical potential $%
\mu $.

We, at first, will \ apply a statistical approach 
\cite{FSV00,VSFC02}.\ This essentially
consists of replacing the single particle density matrix $|n\rangle \langle
n|$ by its value averaged over the energy shell \cite{VSFC02}.

\begin{equation}  \label{form4}
\ \hat{\rho}_{\varepsilon _{n}}=\frac{1}{g(\varepsilon _{n})}\sum_{n^{\prime
}}\delta (\varepsilon _{n}-\varepsilon _{n^{\prime }})|n^{\prime }\rangle
\langle n^{\prime }|=\frac{1}{g(\varepsilon _{n})}\delta (\varepsilon
_{n}-h).
\end{equation}

An asymptotic expression for $\ \hat{\rho}_{\varepsilon_{n} \text{ }}$ can then
be derived using the semi-classical method by Balian-Bloch for infinite 
hard
wall potentials \cite{BB70} or the Thomas-Fermi (TF) or equivalently
Strutinsky averaging method for smooth potentials \cite{RS80}.\ Recognising
that the two body wavefunctions $\langle r_{1}r_{2}|n\overline{n}\rangle $
in the pairing matrix elements can be written as
$\langle r_{1}r_{2}|n\overline{n}
\rangle =$ $\langle r_{1}|n\rangle \langle n| r_{2}\rangle $, we
can pass to the continuum limit and write for (\ref{form1})
\begin{equation}  \label{form4a} 
\Delta(\epsilon) = - \int d\epsilon^{\prime} g(\epsilon^{\prime})
v(\epsilon,\epsilon^{\prime}) \Delta(\epsilon^{\prime})/2 
E(\epsilon^{\prime}).
\end{equation}        
The averaged pairing matrix element is given by
\begin{equation}
\label{form6}      
v(\epsilon,\epsilon^{\prime}) = \int \int d\Gamma d\Gamma^{\prime} 
f f^{\prime} v \left( \mbox{\boldmath$p$} - \mbox{\boldmath$p'$} \right)
\delta( \mbox{\boldmath$R$} -\mbox{\boldmath$R'$}).
\end{equation}       
where
$d\Gamma= d \mbox{\boldmath$R$}d \mbox{\boldmath$p$}/(2 \pi)^3$ 
 and v$\left( \mbox{\boldmath$p$}\right) $ is the
Fourier transform of the pairing force, $f=f_{\epsilon }\left(
\mbox{\boldmath$R,p$}\right) $ is the Wigner transform \cite{RS80} of
$\hat{\rho}_{\varepsilon \text{ \ }}$ in (\ref{form4})
and a prime on $\Gamma$ and $f$ means that all variables should be 
replaced by primed ones. The size dependence
of the gap parameter $\Delta = \Delta\left( \epsilon =\mu \right) $ is then
contained in the corrections to the bulk values of g$\left( \epsilon \right)
$, v$\left( \epsilon ,\epsilon ^{\prime }\right), $ and $\mu $.

Let us first evaluate $\Delta$ for the case of metallic 
grains and
films.\ The electrons be confined by an infinite hard wall potential of
arbitrary shape.\ As usual in condensed matter physics, we approximate the
attractive electron-electron interaction by a delta function
pseudo-potential with a cut-off in energy symmetrically on both sides of the
Fermi energy $\mu $ of the order of the Debye frequency $\omega_{D}$.\ In the bulk the
pairing matrix element is therefore given by $\langle k-k|$v$|k^{\prime
}-k^{\prime }\rangle = \frac{-v_0}{V}$ for$\ \left| \epsilon _{k}-\mu 
\right| $, $\left|
\epsilon _{k^{\prime }}-\mu \right| \leq \omega _{D}$ and zero otherwise 
and $V$ is the volume of the system.
For a finite size grain our main task will be to evaluate the pairing matrix
elements (\ref{form6}) for this case. The expression of the level density
g$%
\left( \epsilon \right) $ in terms of  volume, surface, and curvature
contributions is
well known since long \cite{BB70}. For the matrix elements we also will
employ the Balian-Bloch method \cite{BB70} \ using the method of images. To
lowest order the distribution functions in (\ref{form6}) are given by $%
f_{\epsilon}\left( \mbox{\boldmath$R$},\mbox{\boldmath$p$}\right) \propto 
\delta \left( \epsilon - \hbar^{2}p^{2}/2m\right) $ which is the 
bulk
expression. In order to obtain the correction term, we transform  back into
coordinate representation, $f_{\epsilon}\left( \mbox{\boldmath$R$},\mbox{\boldmath$p$}\right)
\longrightarrow \rho _{\epsilon}\left( \mbox{\boldmath$r$},\mbox{\boldmath$r'$}
\right) $, and then replace $z'$ by $-z'$, the $z$-direction being the one
perpendicular to the surface.\ Back into phase space one obtains,
$f_{\epsilon}(\mbox{\boldmath$R$},\mbox{\boldmath$p$}) =
g(\epsilon)^{-1} \big[ \delta \left( \epsilon - {\hbar 
^{2}p^{2}}/{2m}\right) + \delta f \big] $ with
\begin{equation}
\delta f = - \delta(p_z) \frac{2m/\hbar^2}{k_ {\epsilon}(p_x,p_y)}
cos \big( 2R k_{\epsilon}(p_x,p_y) \big)
\label{form7} 
\end{equation} 
where $k_{\epsilon }\left( p_{x},p_{y}\right) =\left( \frac{2m}{\hbar 
^{2}}%
\left( \epsilon -\frac{\hbar ^{2}}{2m}\left( p_{x}^{2}+p_{y}^{2}\right)
\right) \right) ^{\frac{1}{2}}$.\ Since $f_{\epsilon }\left( %
\mbox{\boldmath$R$},\mbox{\boldmath$p$}\right) $ is normalized to unity, 
one obtains from (\ref{form7}), in integrating over phase space, the 
classical
result for the level density $g\left( \epsilon \right) =\frac{1}{4\pi ^{2}%
}\left( \frac{2m}{\hbar ^{2}}\right) ^{\frac{3}{2}}\sqrt{\epsilon
}V-\frac{S%
}{16\pi }\left( \frac{2m}{\hbar ^{2}}\right)$\cite{BB70}. An important point
to be realised is that the volume V and surface S correspond to the borders
of the hard wall.\ Since the density is diffuse at the surface, the relevant
matter volume $V_{M}< V$ is therefore given by the wall delimitation which
encloses the correct number of particles.\ The relations between V, S and V$%
_{M},$ S$_{M}$ are worked out in \cite{SF83} and are to lowest order given by
V$=$V$_{M}+\frac{3\pi }{8k_{F}}$S$_{M}+...$ and S$=$S$_{M}+....$. The level
density at the Fermi energy then becomes :
\begin{equation}
\label{form8}
g_F=g(\epsilon = \mu)=\frac{V_M}{4\pi ^{2}} \frac{2m}{\hbar^{2}} k_{F }
(1+ \frac{\pi}{8k_{F}} \frac{S_{M}}{V_{M}}+\dots).
\end{equation}
We remark that the sign of the surface term is now positive, that is, for a
given volume V$_{M}$ the level density is {\it enhanced} by the presence of
a diffuse surface which, in fact, is the usual situation.
 With (\ref{form7}) and the definition of $g(\epsilon)$
it is, in considering that $(\delta f)^2$ also contributes 
to order $\frac{S_M}{V_M}$, straightforward to
evaluate the pairing matrix-element (\ref{form6}).\ In the case of our delta
force, its Fourier transform is a constant and one obtains,

\begin{eqnarray}  \label{form9}
&& \text{v} ( \epsilon ,\epsilon ^{\prime }) =
\ \frac{-v_0}{V} 
\left(1 + \frac{\pi}{4} \frac{min(k_{\epsilon},k_{\epsilon^{\prime}})}
{k_{\epsilon} k_{\epsilon^{\prime}}}  
\frac{S}{V}+..\right) \nonumber \\
&=&  \ \frac{-v_0}{V_{M}}
\left(1 + \frac{\pi}{4} \frac{min(k_{\epsilon},k_{\epsilon^{\prime}})}
{k_{\epsilon} k_{\epsilon^{\prime}}}
\frac{S_M}{V_M} - \frac{3 \pi}{8 k_F} \frac{S_{M}}{V_{M}}+..\right).
\end{eqnarray}
 We therefore see that, contrary to the level density, the matrix 
element $v_F=\text{v}(\mu,\mu)$ diminishes in absolute size in the 
presence of a surface. All ingredients are now prepared 
and one can solve the gap equation (\ref{form4a}) for instance 
numerically. However, 
there exists a well known and accurate analytical solution which is more 
interesting \cite{FW71}.
The result is $\Delta =2\omega _{D}\exp \left(\frac{1}{v_F g_F}\right)$ 
. Inserting $g_F$ from
(\ref{form8}) and $v_F$ from (\ref{form9}) into the above expression, we 
notice that the product $v_F g_F$ does not depend on the surface.
 However, one also has to account for the 
compression effect due to the surface tension which increases the chemical 
potential or respectively the Fermi momentum, and thus $g_F$.
 Finally this leads to an enhancement of the gap for low system sizes.
 Elaborating one obtains $k_{F}=k_{F}^{B}\left( 1+\frac{\pi
}{8}\frac{1}{k_{F}^{B}}\frac{S_{M}}{V_{M}}\right)$, where $k_F^B$ stands 
for the bulk value. Inserting into the expression for the gap one obtains
\begin{equation}  \label{form10}
\Delta =\Delta _{B}e^{- \frac{1}{v_{F}^{B} g_{F}^{B}}\frac{\pi 
}{8}\frac{1}{k_{F}^{B}}%
\frac{S_{M}}{V_{M}}},  \label{formula}
\end{equation}
where $v_F^B$ and $\Delta_B$ stand for bulk values. One clearly sees 
that the 
gap becomes enhancend as the size of the system {\it decreases}.          
     
%
It is fortunate that formula (\ref{form10}) can be tested on a 
very early quantum mechanical
solution of (\ref{form1}) for a slab \cite{TB63a}. 
In this case one has
$\frac{S_{M}}{V_{M}}=\frac{2}{L}$ where $L$ is the film thickness.
 In \cite{TB63a} the constants in (\ref{form10}) were chosen 
$-v_F^B g_{F}^{B}=0.3$ and $k_{F}^{B}=0.84 \times 10^{8} cm^{-1}$.
It can be seen from Fig.1 that our pocket formula passes on average well 
through the quantum mechanical values \cite{TB63a}.

\begin{figure}[h]
\centering
\includegraphics[height=8cm,angle=270]{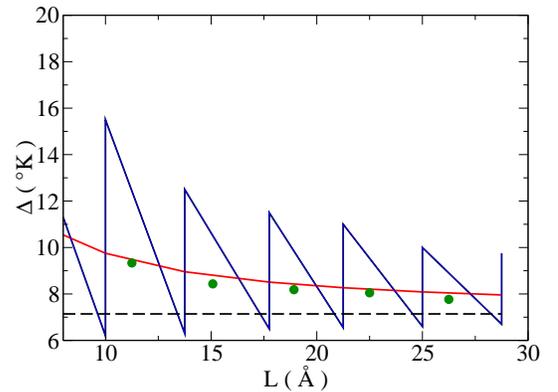}
\caption{\label{fig1}
Dependence of the gap, for the case of a
superconducting homogeneous film, on the film thickness L. The saw
tooth line corresponds to a quantum mechanical calculation \cite{TB63a}, 
whereas the
smooth curve corresponds to formula (8).  
The horizontal line represents the bulk value $\Delta_B$ for Aluminium.
The dots represent the center of gravity of the triangles in which they 
are lying (a crude way to estimate an average of the quantal results)}
\end{figure}
 In refs. \cite{BR96,RB97,CA68,DFGGO73} it is indicated that in
the case of Al grains one obtains, with respect to the bulk, an enhancement for the critical
temperature T$_{c}$ by roughly a factor of two for a grain diameter of 45\AA
.\ For a spherical grain with $V_{M}=\frac{4 \pi R^3}{3}$ one obtains
$\frac{S_{M}}{V_{M}}=\frac{3}{R}$.   
 However, grains are rather pancake shaped than spherical 
\cite{BR96,ZG69}. For an oblate ellipsoid with short diameter half the 
one of a sphere with the same volume the increase of $S_M/V_M$ is 44 
\%. Probably grains are even triaxial (see ref.\cite{ZG69}, fig.2) 
 and we take $S_M/V_M=9/(2R)$ which corresponds to a 50 \% 
increase over the spherical case. Taking in (\ref{form10}) the bulk 
values for Al 
 that is $k_F^B = 1.75 \AA^{-1}$  and $-v_F^B g_F^B$=0.168, we 
obtain 
from  (\ref{form10}) for $\Delta/\Delta_B$ an enhancement $\sim$ 30 \% at 
$2R \sim 45 \AA$ which is a sizeable fraction of the experimental value. 
However, in such small grains the electron 
levels are discrete and it is well known \cite{RS80}  that the gap 
equation has no solution, if the average level distance $d \gg \Delta_B$. We 
therefore solved the gap equation (\ref{form1}) for the picked fence model 
(equally spaced levels with Kramers degeneracy) \cite{BD99}
 for $\omega_D$ 
= 395 K which is the value for Al. The number of levels $n_W$ in the 
window 
2$\omega_D$ was estimated to be i) $n_W^B = 2 \omega_D g^B_F$ if we take 
only the lowest order term in (\ref{form8}) and ii) $n_W = 2 \omega_D 
g_F$ when including the surface correction to the level density (and the 
one coming from $\mu$, see above). 
For the dimensionless interaction constant we take $- \lambda \equiv v_F 
g_F = v_F^B g_F^B \left( 1+\frac{\pi}{8}\frac{1}{k_{F}^{B}}\frac{S_{M}}{V_{M}}
\right)$, with $v_F^B g_F^B$ as above. In this way we also can calculate 
$\Delta/\Delta_B$ quantally in the picket fence model. We find that 
$\Delta/\Delta_B$ raises from $\Delta/\Delta_B=1$ for $R = \infty$ to 
$\Delta/\Delta_B \sim 1.2$ at $2R \sim 60 \AA$, following quite accurately 
our pocket formula. For smaller grain sizes the solution of the gap 
equation quickly breaks down, the critical size occurring at $2R_c 
\simeq 40 \AA$. The situation is summarized in Table 1. It therefore 
seems within our schematic model that one can only reach a moderate 
enhancement of 20\% - 30\% depending on whether or not one believes 
into a continuation of the increase into the pair-fluctuating regime.
\begin{table}
\begin{center} \small
\begin{tabular}{l c c c c c}
$n_W$ & $2 \tilde{R} [\AA]$ & $2 R [\AA]$ & $\tilde{\Delta} [K]$ & 
$\Delta [K]$ & eq.(\ref{form10}) [K]\\
\hline
  60 &  41.49 &  40.83 & 0.00 & 0.00 & 1.34 \\
  80 &  45.73 &  45.06 & 0.00 & 0.00 & 1.31 \\
 100 &  49.30 &  48.64 & 0.00 & 0.83 & 1.28 \\
 200 &  62.22 &  61.55 & 0.95 & 1.18 & 1.22 \\
 300 &  71.26 &  70.60 & 1.00 & 1.18 & 1.19 \\
 400 &  78.46 &  77.79 & 1.00 & 1.16 & 1.17 \\
 500 &  84.53 &  83.86 & 1.00 & 1.15 & 1.15 \\
1000 & 106.54 & 105.87 & 1.00 & 1.12 & 1.12 \\
\hline
\end{tabular}
\end{center}
\caption{Number of levels in the window ($n_W$), size 
($2\tilde{R}$), ($2R$) and gap ($\tilde{\Delta}$), ($\Delta$) {\bf 
without} 
and {\bf with} surface correction, respectively. The gap obtained using 
eq.(\ref{form10}) is also given.}  
\end{table}
Several comments are, however, in order: Equal level spacing is the most 
unfavorable situation which can exist . Usually a certain percentage 
of grains have some symmetries which can enhance the gap (see ref. 
\cite{GFD01}). Therefore on average the gap is larger than the one we 
have calculated and correspondingly $R_c$ is smaller. However, a 
precise estimate of the effect is difficult. The gap can also be 
calculated from the exact solution of the picket fence model 
(see \cite{DS99}). It turns out that this 'quantal' 
definition of the gaps 
yields, around the phase transition region, substantially larger values  
than those from the mean field BCS 
theory, again enhancing the ratio $\Delta/\Delta_B$. The quantal 
values of $\Delta$ also can be obtained for sizes quite a bit 
smaller than $R=R_c$ of BCS theory. We therefore think to have isolated 
an important enhancement mechanism of pairing in metallic nano-grains, 
stemming from the presence of a surface.
Other effects, like e.g. the size dependence of the phonon spectrum, 
should be taken into account to obtain quantitative agreement with 
experimental data.

\begin{figure}[h]
\centering
\includegraphics[height=8cm,angle=270]{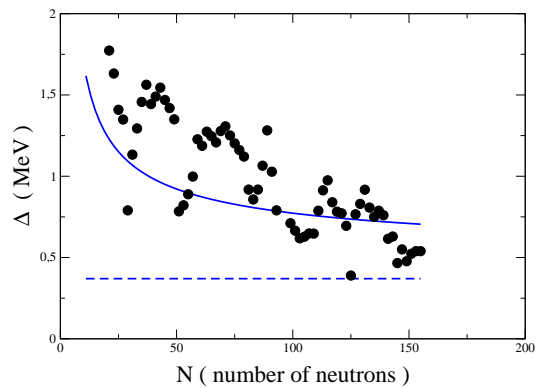} \label{fig2}
\caption{Average nuclear gaps as a function of neutron number N along the
valley of $\beta$-stability of the nuclear chart. The experimental points
have been taken from \cite{HBGSS02}. Broken line: the asymptotic value 
$\Delta_B$=0.37 MeV to which the full line converges.}
\end{figure}

In nuclear physics it is well known since decades that pairing is stronger
in lighter nuclei than in heavier ones. An empirical formula $\Delta =12/
\sqrt{A}$ with $A=N+Z$ the sum of neutron ($N$) and proton ($Z$) numbers had been
used in the past to fit the data \cite{BM69,RS80}. However more recently 
Satula et al.
\cite{SDN98} pointed out that the data used so far to extract the gap values
were overestimated\ and contaminated by the Jahn-Teller effect \cite{SDN98}. A
new analysis using the filter  $\Delta =\frac{1}{2}\left[
E_{0}^{N+1}+E_{0}^{N-1}-2E_{0}^{N}\right] $ for neutron number $N$ odd 
only, $E_{0}^{N}$ being the measured binding energies of
nuclei, revealed that the mass number dependence of $ \Delta $ is
substantially weaker than the $12/\sqrt{A}-$law. In nuclear physics
it is common use to solve the gap equation (1) either, as for the metallic
grains, also  using a $\delta $-force pseudo-potential with a cut-off
\cite{BM69,RS80} or more sophisticated finite range forces are employed for the
matrix elements in (1) not necessitating any cut-off. One of the best tested
and successful forces of the latter type is the Gogny D1S force \cite{BGG91}.
In principle for nuclei it is more appropriate to work with
smooth potentials like the Woods Saxon or harmonic oscillator potentials and
to use for the average density matrix on the energy shell (3) the well known
Wigner-Kirkwood $\hbar$-expansion \cite{RS80}. This procedure is, however, more
cumbersome and does not lead to such a handy formula as (\ref{form10}). For
space reason we cannot present this here and it will be published separately
in the future. For the time being we will also use (\ref{form10}) for finite
nuclei as a generic formula. In nuclear physics the convention is such 
that $-v_F^B=\frac{v_0}{V_M} = G$ and $g_F^B = 
\frac{1}{4}\frac{6}{\pi^2}a$ where the level density parameter $a = 
 \frac{\pi^2}{4} \frac{2 m^*}{\hbar^2 {k_F^B}^2}A$ MeV$^{-1}$. 
An average value from Skyrme and Gogny forces is 
$a \sim \frac{A}{20}$ 
MeV$^{-1}$. A typical value for $G$ which can be found in the literature 
\cite{RS80,EG76} is $G=\frac{25}{A}$ MeV. We also checked, using the 
methods of \cite{VSFC02}, that this latter value is compatible with the 
Gogny D1S force \cite{BGG91}.

  
On average nuclei are spherical
and then $\frac{S_{M}}{V_{M}}=\frac{3}{R}$ where $R=r_{0}A^{\frac{1}{3}}$ is
the nuclear radius. The product k$_{F}^{B}r_{0}=(\frac{9\pi}{8})^{\frac{1}{3}}$
 is a universal
number and then, besides $\Delta _{B}$, all constants in (\ref{form10}) are
fixed also for the nuclear case. 
The bulk value of the gap is a quantity which in nuclear physics is quite
uncertain because the mass number range of nuclei is too small to
 extrapolate to infinite nuclear matter without the guidance of a
reliable formula. We expect (\ref{form10}) to be such an expression which
allows to pin down $\Delta_{B}$ within certain limits. In Fig.2 \ref{fig2}
we show that a good fit to the data with the above values for $a$ and 
$G$ is obtained with $\Delta _{B}=.37$
MeV. Using for $a = \frac{A}{16}$MeV$^{-1}$ which is obtained with 
$m=m^*$ and which is the standard Fermi gas value used in 
 phenomenological models, the fit yields $\Delta_B=0.45$. 
This gives a slightly flatter but still acceptable curve than the one 
shown in Fig.2 and shows that formula (\ref{form10}), for the nuclear 
case, is quite robust. These values for $\Delta_B$ are of the same order 
of magnitude 
as the asymptotic value $\Delta_B=0.58 $ MeV calculated from the D1S force 
\cite{BGG91}. 
In Fig.\ 2 the $A$-dependence has been converted into an
$N$-dependence via the relation $A-N=A/(1.98+0.0155A^{\frac{2}{3}})$ which
defines the valley of stability of the nuclear chart
\cite{MS69}. 
Therefore
for nuclei the pocket formula (\ref{form10}) gives a very satisfying
reproduction of the data and we thus conclude that it contains the
essentials of the physics. 

In conclusion, we isolated in this work an important and generic 
enhancement factor of pairing in finite Fermi systems. This stems from 
the surface corrections to their respective bulk values of level density, 
 pairing matrix element, and chemical potential. 
We derived a pocket formula for the enhancement factor $\Delta/\Delta_B$ 
which is very general and depends exponentially on the ratio surface to 
voulume of systems of arbitrary shape. It 
 remains valid for level spacings $d \le 1.4 \Delta$ because 
for larger spacings the solution of the gap equation breaks down. 
Our theory explains satisfactorily the average experimental mass number 
dependence of nuclei. For Al grains we obtain within the picket fence 
model a maximun enhancement of $\Delta/\Delta_B \sim$ 1.2 at a grain 
diameter of $\sim$ 6 nm. We checked that the situation is similar for the 
case of Sn grains  \cite{ZG69}.
This estimate is based on BCS theory. 
We, however, argue that in a more 
realistic theory the corresponding gap may exist for smaller grains 
yielding a more important fraction of the experimental results. 
 Other effects mentioned above can give additional 
enhancements. Studies in this direction are planned for the future.

\begin{acknowledgements}
Acknowledgements: We gratefully acknowledge extended discussions and 
informations with G. Deutscher. We also appreciated interest and 
discussions  with J-F. Berger, O.Bohigas,  O. Buisson, J. Dukelsky, M. 
Girod, S. Hilaire, P. Leboeuf, P. Nozi\`eres, 
N. Pavloff, J. Pekola  and W. Satula. X.V. acknowledges financial support 
from DGCYT 
(Spain) under grant PB98-1247 and from DGR (Catalonia) under grant 
2001SGR00064, F.H. is supported by Institut Universitaire de France.
 \end{acknowledgements}

\bibliographystyle{abbrv}
\bibliography{bib/physical}
\end{document}